\documentclass{llncs}
\usepackage{latexsym,amsmath, algorithm2e}
\usepackage{graphicx}
\usepackage{verbatim}

\begin{document}

\bibliographystyle{plain} 

\title{Bounded Search Tree Algorithms for Parameterized Cograph Deletion: Efficient Branching Rules by Exploiting Structures of Special Graph Classes\thanks{Supported in part by NSERC Discovery Grant RGPIN 327587-09}}
\author{James Nastos \and Yong Gao \thanks{We thank Dr. Donovan Hare for our discussions on these results}}
\institute{
    Department of Computer Science, Irving K. Barber School of Arts and Sciences.\\
    University of British Columbia Okanagan, Kelowna, Canada V1V 1V7\\
    \email{jnastos@interchange.ubc.ca, yong.gao@ubc.ca}
}

\maketitle

\begin{abstract}
Many \emph{fixed-parameter tractable} algorithms using a bounded search tree have been repeatedly improved, often by describing a larger number of branching rules involving an increasingly complex case analysis. We introduce a novel and general search strategy that branches on the forbidden subgraphs of a graph class relaxation. By using the class of $P_4$-sparse graphs as the relaxed graph class, we obtain efficient bounded-search tree algorithms for several parameterized deletion problems. We give the first non-trivial bounded search tree algorithms for the cograph edge-deletion problem and the trivially perfect edge-deletion problems. For the cograph vertex deletion problem, a refined analysis of the runtime of our simple bounded search algorithm gives a faster exponential factor than those algorithms designed with the help of complicated case distinctions and non-trivial running time analysis~\cite{NiRo} and computer-aided branching rules~\cite{GGHN}.

\smallskip
\textbf{Keywords:} Fixed-parameter tractability; edge-deletion; graph modification; cographs; trivially perfect graphs; quasi-threshold graphs; bounded search tree.

\end{abstract}

\normalsize

\section{Introduction}

A graph is a \emph{cograph}~\cite{Sein} if it has no induced subgraph isomorphic to a $P_4$, an induced path on four vertices. The name originates from \emph{complement reducible graphs} as cographs are also characterized as being those graphs $G$ which are either disconnected or else its complement $\overline{G}$ is disconnected~\cite{Sein}. They are a well-studied class of graphs and many NP-complete problems on graphs have been shown to have polynomial time solutions when the input is a cograph~\cite{CPS}.

A \emph{graph modification problem} is a general term for a problem that takes a graph as input and asks how the graph can be modified to arrive at a new graph with a desired property. Usually, graph modifications are edge additions or deletions, vertex additions or deletions, or combinations of these. Our work on the following problems originally stems from studying social networks from which edge removals are made to reveal underlying structures in the network.

Many parameterized graph modification problems are tackled by a bounded search tree method where the size of the search tree usually dominates the computation time. This paper presents a framework for designing branching rules for bounded search tree algorithms. Our strategy exploits the structure of specialized graph classes in order to design efficient algorithms for (hard) problems on general graphs. We illustrate this method by giving algorithms that solve four different graph modification problems.

The \emph{cograph edge-deletion problem} is the problem of determining when a graph $G=(V,E)$ has a set $S$ of at most $k$ edges which can be removed in order to make $G_2=(V,E \setminus S)$ a cograph. This problem is known to be NP-complete \cite{ElMC} and also known to be \emph{fixed parameter tractable}~\cite{CAI}. Similarly, the \emph{trivially perfect edge-deletion problem} asks whether $k$ edges can be removed to turn a graph into a trivially perfect graph (a graph which is $P_4$ and $C_4$-free.) We first show how to solve these problems in linear time on a relaxed graph class, the \emph{$P_4$-sparse} graphs, and design algorithms to solve these problems on general graphs by branching towards $P_4$-sparse graphs. Furthermore, we give improved algorithms for the vertex-deletion version of these problems.

We note that since the class of cographs is self-complementary, an algorithm solving $k$-edge-deletion problem also serves as a solution to the problem of $k$-edge-addition to cographs. Similarly, the $k$-edge-deletion problem to trivially perfect graphs serves as a solution to the $k$-edge-addition problem to co-trivially perfect graphs.

This paper is structured as follows: Section 2 summarizes previous results related to graph modification problems and gives some background on the class of $P_4$-sparse graphs; Section 3 gives edge-deletion algorithms to cographs and to trivially perfect graphs; Section 4 designs \emph{vertex deletion} algorithms for cographs and \emph{trivially perfect} graph and their improvements using {\sc Hitting Set}; Section 5 summarizes and discusses these results and suggestions a number of directions for future work.

\section{Previous Results and Background}

\subsection{Previous Fixed-Parameter Tractability Results}

While cographs can be recognized in linear time~\cite{CPS}, it is also known that it is NP-complete to decide whether a graph is a cograph with $k$ extra edges~\cite{ElMC}. Graph modification problems have been studied extensively: Yannakakis shows that vertex-deletion problems to many types of structures is NP-hard~\cite{Yanna}. Elmallah and Colbourn give hardness results for many edge-deletion problems~\cite{ElMC}.

Recently, much research has been devoted to finding fixed-parameter tractable algorithms for graph modification problems: Guo~\cite{Guo} studied edge deletion to split graphs, chain graphs, threshold graphs and co-trivially perfect graphs; Kaplan et al.~\cite{KST} studied edge-addition problems to chordal graphs, strongly chordal graphs and proper interval graphs; Cai~\cite{CAI} showed fixed-parameter tractability for the edge deletion, edge addition, and edge editing problem to any class of graphs defined by a finite set of forbidden induced subgraphs. The constructive proof implies that $k$-edge-deletion problems to a class of graphs defined by a finite number of forbidden subgraphs is $O(M^kp(m+n))$ where $p$ is some polynomial and $M$ is the maximum over the number of edges in each of the forbidden induced subgraphs defining that graph class in question. For $k$-edge-deletions to $P_4$-free graphs in particular, Cai's result implies an algorithm running in $O(3^k(m+n))$ time. This algorithm would work by finding a $P_4$: $a-b-c-d$ in a graph and branching on the 3 possible ways of removing an edge in order to destroy the $P_4$ (that is, removing either the edge $\{a,b\}$ or $\{b,c\}$ or $\{c,d\}$).

Nikolopoulos and Palios study the edge-deletion to cograph problem for a graph $G-xy$ where $G$ is a cograph and $xy$ is some edge of $G$~\cite{NikoPalio}. Lokshtanov et al. study cograph edge-deletion sets to determine whether they are \emph{minimal}, but not a minimum edge-deletion set~\cite{LMP}. To the best of our knowledge, ours is the first study that specifically addresses the edge-deletion problem to cographs. We present a bounded search tree algorithm that solves $k$-edge-deletion to cographs in $O(2.562^k(m+n))$ time by performing a search until we arrive at a $P_4$-sparse graph and then optimally solving the remainder of the problem using the structure of $P_4$-sparse graphs.

Graph modification problems can also be regarded as a type of graph recognition problem. Following the notation of Cai~\cite{CAI}, for any class of graphs $\mathcal{C}$, we call $\mathcal{C}+ke$ the set of all graphs which cane be composed by adding $k$ extra edges to a graph from class $\mathcal{C}$. Similarly, $\mathcal{C}-ke$ is the set of graphs which are formed from a graph from class $\mathcal{C}$ with $k$ edge removals. Replacing `edges' by `vertices' in these definitions gives analogous classes for $\mathcal{C}+kv$ and $\mathcal{C}-kv$. A $k$-edge-deletion problem to a class of graphs $\mathcal{C}$ can thusly be restated as a recognition problem for the class of $\mathcal{C}+ke$ graphs. Our results on cographs here can be restated as recognition algorithms for the classes: Cograph+$ke$, Cograph-$ke$, Cograph+$kv$, Trivially Perfect+$ke$.

\subsection{Background Information: $P_4$-sparse graphs}

\begin{figure}
  \includegraphics[height=8cm]{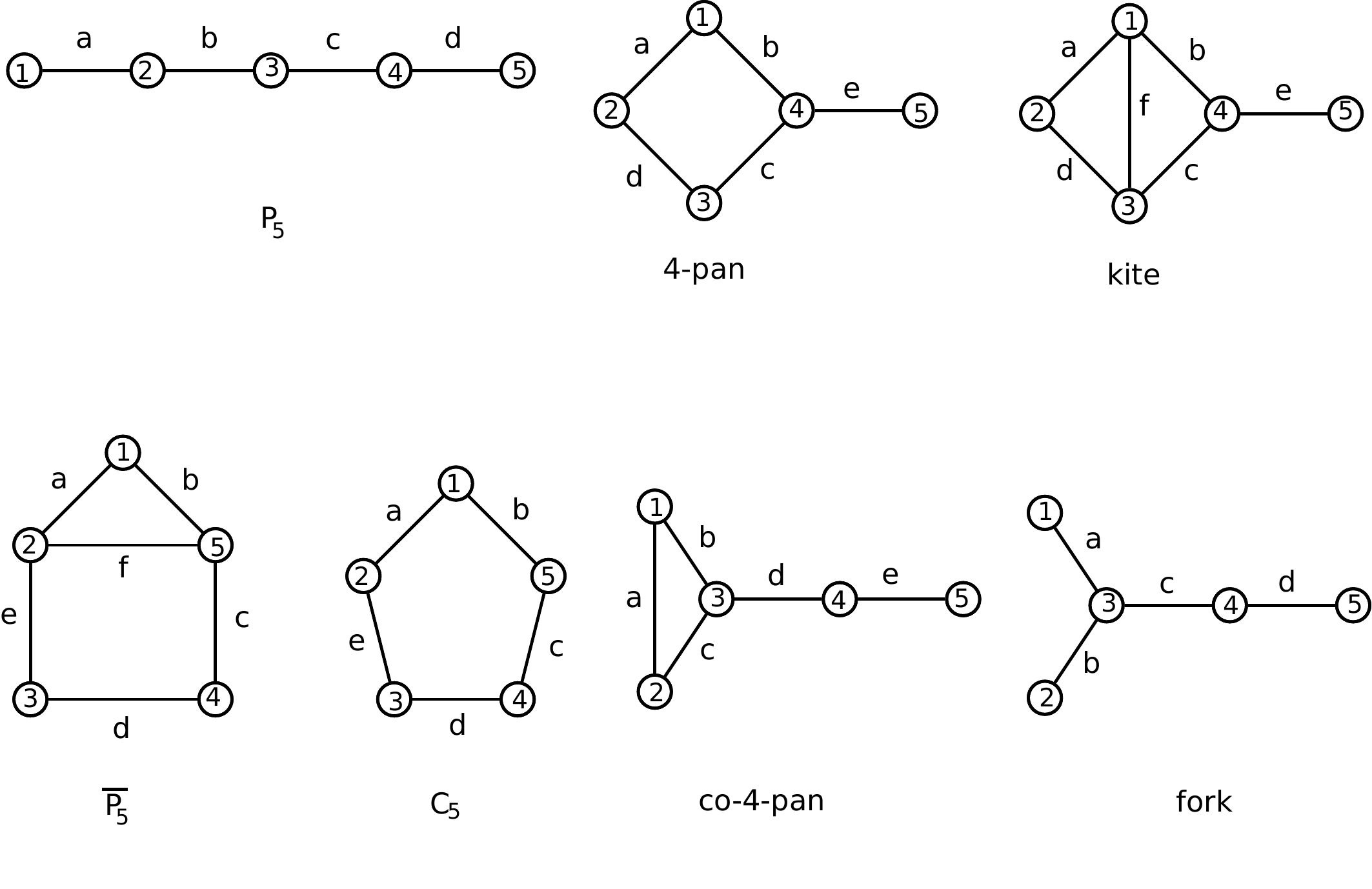}\\
  \caption{The forbidden induced subgraphs for $P_4$-sparse graphs}\label{forb}
\end{figure}

One generalization to the class of cographs is formed by allowing $P_4s$ to exist in a graph but in restricted amounts. Ho\'ang \cite{Hoa} introduced $P_4$-sparse graphs to be those for which every induced subgraph on five vertices induces at most one $P_4$. This immediately implies a forbidden induced subgraph characterization which restricts any subgraph of five vertices inducing two or more $P_4$s. We include these graphs in Figure 1.

A special graph structure called a \emph{spider}~\cite{JaOl} commonly occurs in graph classes of bounded cliquewidth. We define two types of spiders here:

\begin{definition}
A graph $G = (V,E)$ is a \emph{thin spider} if $V$ can be partitioned into $K$, $S$ and $R$ such that:
\begin{itemize}
\item[i)] $K$ is a clique, $S$ is a stable set, and $|K| = |S| \geq 2$.
\item[ii)] every vertex in $R$ is adjacent to every vertex of $K$ and to no vertex in $S$.
\item[iii)] each vertex in $S$ has a unique neighbour in $K$, that is: there exists a bijection $f:S \rightarrow K$ such that every vertex $k \in K$ is adjacent to $f(k) \in S$ and to no other vertex in $S$.
\end{itemize}
\end{definition}

\begin{figure}
    \hspace{2.6cm}
  \includegraphics[height=5cm]{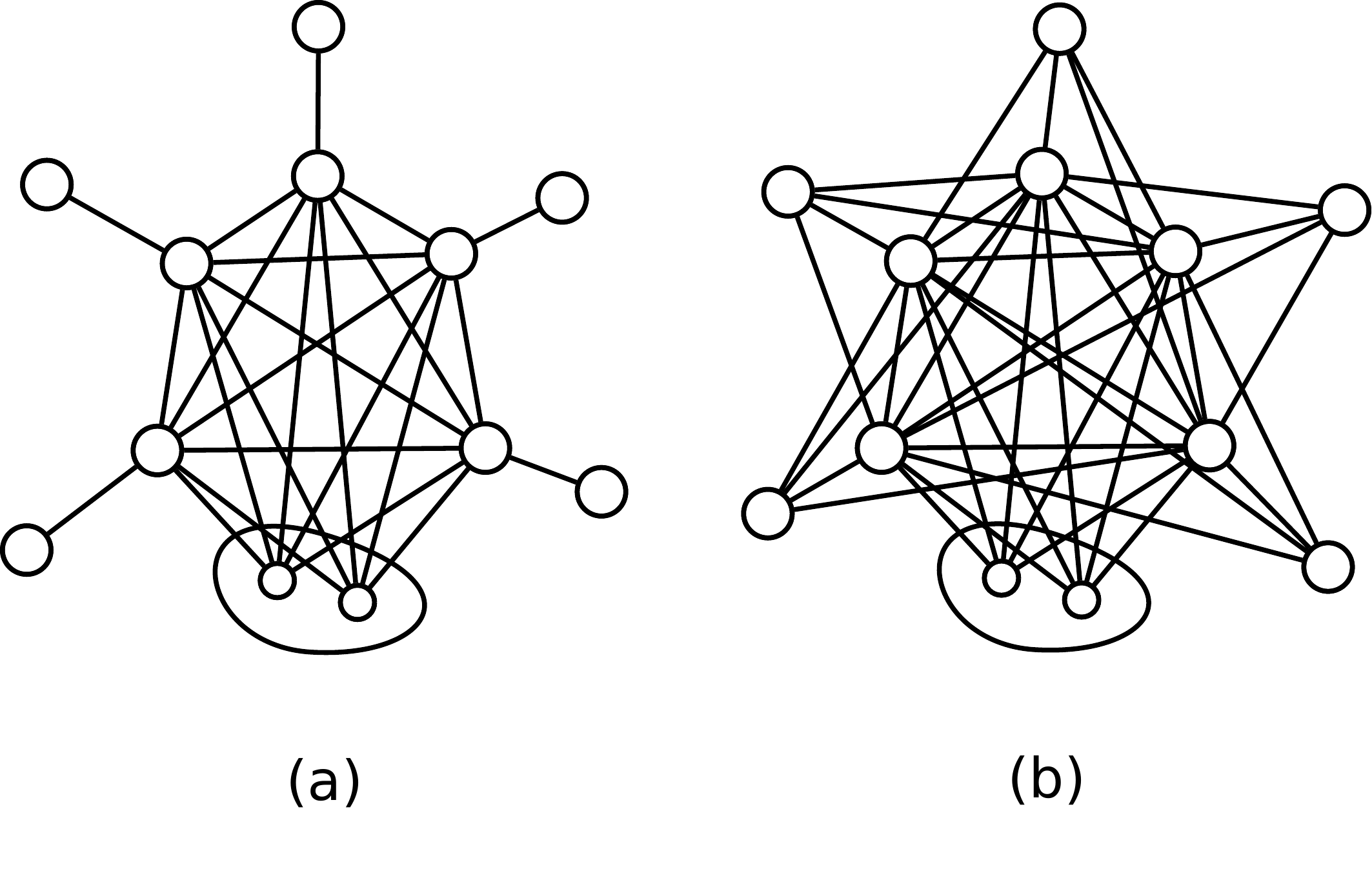}\\
  \caption{(a) A thin spider and (b) thick spider with $|K| = |S| = 5$ and $|R|=2$ }\label{spider}
\end{figure}

A graph $G$ is called a \emph{thick spider} if $\overline{G}$ is a thin spider. Note that the vertex sets $K$ and $S$ swap roles under graph complementation, that condition (i) and (ii) hold for thick spiders, and that statement (iii) changes to saying that every vertex in $S$ has a unique \emph{non-neighbour} in $K$. The sets $K$, $S$ and $R$ are called the \emph{body}, \emph{feet} and \emph{head} of the spider, respectively. The edges with one endpoint in $S$ are called \emph{thin legs} or \emph{thick legs} for thin spiders or thick spiders, respectively. Examples of spiders are given in Figure~\ref{spider}.

Ho\`ang \cite{Hoa} defined a graph $G$ to be $P_4$-sparse if every induced subgraph with exactly five vertices contains at most one $P_4$. The following decomposition theorem for $P_4$-sparse graphs was proven in~\cite{JaOl}:
\begin{lemma}\cite{JaOl}
Let $G$ be a $P_4$-sparse graph. Then exactly one of the following is true:
\begin{itemize}
\item[i)] $G$ is disconnected
\item[ii)] $\overline{G}$ is disconnected
\item[iii)] $G$ is a spider
\end{itemize}
\end{lemma}

We conclude our discussion of $P_4$-sparse graphs with an observation that will be useful to us:

\begin{lemma}\label{lem:unique}
  Let $G$ be a spider with body $K$ and feet $S$. Then every edge $\{k_1,k_2\}$ with $k_1, k_2 \in K$ is in one $P_4$ in $K\cup S$. 
\end{lemma}

\begin{proof}
  A $P_4$ can not contain 3 vertices of $K$. If $G$ is a thin spider, let each $k_i$ be adjacent to each $s_i$. The edge $\{k_1,k_2\}$ is only in the $P_4 \ \{s_1,k_1,k_2,s_2\}$. If $G$ is a thick spider, let each $k_i$ be adjacent to every foot $s_j$ where $i\neq j$. The edge $\{k_1,k_2\}$ is only in the $P_4 \ \{s_1,k_2,k_1,s_2\}$.
 \hfill $\Box$
\end{proof}

\section{Edge-Deletion Algorithms}

In this section, we give algorithms for two edge-deletion problems.

\begin{problem}{\sc Cograph Deletion} $(G, k)$:\\
Given graph $G=(V,E)$, does there exist a set $S$ of at most $k$ edges such that $(V,E\setminus S)$ is a cograph?
\end{problem}

A graph is \emph{trivially perfect} if it has no induced subgraphs isomorphic to a $P_4$ or a $C_4$~\cite{YCC}. Every trivially perfect graph is a cograph and every cograph is a $P_4$-sparse graph.

\begin{problem}{\sc Trivially Perfect Deletion} $(G, k)$:\\
Given graph $G=(V,E)$, does there exist a set $S$ of at most $k$ edges such that $(V,E\setminus S)$ is a trivially perfect graph?
\end{problem}

The idea of the algorithms in this section is to focus on the forbidden subgraphs of $P_4$-sparse graphs so that efficient branching rules can be designed systematically. This depends critically on whether these problems can be solved polynomially on $P_4$-sparse graphs. We first show how to solve the cograph deletion problem on $P_4$-sparse graphs in linear time.

\subsection{Computing Cograph Edge-Deletion Sets on $P_4$-sparse Graphs in Linear Time}

We show that a linear time divide-and-conquer algorithm can be designed to find the minimum cograph deletion set for $P_4$-sparse graphs.

\begin{definition}
Let $G$ be a graph and $\overline{G}$ be the complement of $G$. The vertex sets corresponding to the maximal connected components of $\overline{G}$ are called the \emph{co-components} of $G$. If $\overline{G}$ is connected, then we say that $G$ is \emph{co-connected.}
\end{definition}

\begin{proposition}\label{prop:decomposition}
 Let $G$ be a $P_4$-sparse graph and $M(G)$ be the size of a minimum edge-deletion set required to turn $G$ into a $P_4$-free graph. Then:
 \begin{itemize}
 \item[i)] if $G$ is disconnected with components $V_1, \ldots , V_t$, then $M(G) = \sum_{i=1}^t M(V_i)$
 \item[ii)] if $\overline{G}$ is disconnected with co-components $V_1, \ldots , V_t$, then $M(G) = \sum_{i=1}^t M(V_i)$
 \item[iii)] if $G$ is a spider with head $R$, body $K$ and feet $S$, then $$M(R\cup K\cup S) = M(R) + M(K\cup S)$$.
 \end{itemize}
\end{proposition}

\begin{proof}
 (i) This follows from the fact that a $P_4$ is connected and so any $P_4$ is in only one connected component, even after some edge deletions.

 (ii) It is easy to verify that an edge joining two vertices in separate co-components can not be in a $P_4$ (or else in the graph complement this would imply a $P_4$ contains vertices in separate connected components as a $P_4$ is self-complementary.) After any edge-deletions within a co-component are made, the vertex sets of separate co-components are still completely joined, and so any new $P_4$s will not include any two vertices in separate co-components.

 (iii) Call a \emph{leg edge} any edge joining a vertex $s \in S$ with a vertex $k \in K$, a \emph{head edge} any edge joining some $r_1 \in R$ with some $r_2 \in R$, a \emph{body edge} any edge joining two vertices in $K$, and call a \emph{neck edge} any edge joining some $r \in R$ with some $k \in K$.
  
 The structural definition of a spider says that every vertex in $K$ is adjacent to every vertex in $K \cup R$, even after the removal of any leg edges and head edges. Thus a $P_4$ can never contain an edge $\{r,k\}$ with $r\in R$ and $k\in K$ even after leg and head edge removals. We will show that there is an optimal solution without body edges.
  
 Consider an edge-deletion set $E'$ such that $G-E'$ is a cograph, and let $E'' \subset E'$ be the set of body edges and neck edges in $E'$. Consider the $P_4$s in $G - E' + E''$ (the $P_4$s created when adding $E''$ back to $G$.) In $G - E' + E''$, $K$ and $R$ are completely joined and $K$ is a clique and so no $P_4$ crosses the neck. So any $P_4$s in $G-E'+E''$ are strictly in $K\cup S$ or strictly in $R$. Since $E'$ is a cograph deletion set, the induced graph on $R$ in $G-E'+E''$ is $P_4$-free. In $K\cup S$, the body edges added back may be in a $P_4$ with two leg edges, and if so, this $P_4$ will be unique by Lemma~\ref{lem:unique}. Adding the body edges from $E''$ can not create a $P_4$ involving a body edge not in $E''$, so we just concentrate on the unique $P_4$ that each of these added body edges may have created. By deleting one of these leg edges for each body edge that creates a $P_4$, we create a new deletion set $E'-E''+ E'''$ where $E'''$ is a set of leg edges and $|E'''| \leq |E''|$, so this new edge deletion set is a solution no larger than $E'$ which does not use body or neck edges.
 \hfill $\Box$
\end{proof}

We note that parts (i) and (ii) of Proposition~\ref{prop:decomposition} apply to any graph $G$, and not just $P_4$-sparse graphs.

\begin{lemma} \label{thin}
  Let $G$ be a thin spider with body $K = \{k_1, \ldots , k_{|K|}\}$ and legs $S = \{s_1,\ldots ,s_{|K|}\}$, and $\{s_i, k_j\}$ is an edge if and only if $i=j$. Then a minimum cograph edge-deletion set for $K \cup S$ is $\{ \{s_i, k_i\}, i=1..|K|-1\}$.
\end{lemma}

\begin{proof}
  Since $K$ is a clique and $S$ is stable, every $P_4$ in $K \cup S$ has its endpoints in $S$. Furthermore, every pair of vertices in $S$ are in a unique $P_4$. Deleting any $|S|-1$ thin legs will clearly destroy all of the $P_4$s, so this edge-deletion set is indeed a cograph edge-deletion set. To see that it is of minimum size, assume there is a deletion set of size $|K|-2$ or less in which two legs are not part of the deletion set. Let these two legs be $\{s_1,k_1\}$ and $\{s_2,k_2\}$ and call them ``permanent" in this case. Since $\{s_1, k_1, k_2, s_2\}$ is a $P_4$ and the edges $\{s_1,k_1\}$ and $\{s_2,k_2\}$ are not in the deletion-set, it must be that $\{k_1, k_2\}$ is in the deletion set. There at most $|K|-3$ other edges in the deletion set. Now $\{s_1, k_1, k_j, k_2\}$ induces a $P_4$ for every $j=3\ldots |K|$. This means that the permanent edge $\{s_1,k_1\}$ is still in $|K|-2$ $P_4$s and every pair of these $P_4$s have distinct edges aside from $\{s_1,k_1\}$. Thus it is impossible to destroy all of these remaining $P_4$s with only $|K|-3$ additional deletions or less.
  \hfill $\Box$
\end{proof}

\begin{lemma} \label{thick}
  Let $G$ be a thick spider with body $K = \{k_1, \ldots , k_{|K|}\}$ and feet $S = \{s_1,\ldots ,s_{|K|}\}$, and $\{s_i, k_j\}$ is an edge if and only if $i\neq  j$. Then a minimum cograph edge-deletion set for $K \cup S$ is $\{ \{k_i, s_j\}, i<j \}$.
\end{lemma}

\begin{proof}
 Every edge in $K \cup S$ is in exactly one $P_4$: an edge $\{k_i,k_j\}$ is only in the $P_4$ $\{s_j, k_i, k_j, s_i\}$ and any edge $\{s_i, k_j\}$ is only in the $P_4$ $\{s_i, k_j, k_i, s_j\}$ so the number of $P_4$s in $K \cup S$ is $\binom{|S|}{2}$, and since no two of these $P_4$s share an edge, at least $\binom{|S|}{2}$ deletions are required. Consider the edge set $T = \{ \{k_i,s_j\}, i<j\}$. When deleting $T$ from $K \cup S$, $K$ is still a clique and $S$ is still a stable set, and so if there is any $P_4$ in $(K \cup S) \setminus T$, its endpoints must still be in $S$. But after deletion of $T$, we have that the neighbourhood of $s_i$ is $N(s_i) = \{k_{i+1}, \ldots ,k_{|K|}\}$ which means that $N(s_i) \subset N(s_j)$ for all $i>j$, and so no two vertices in $S$ can be the endpoints of a $P_4$. So $T$ indeed destroys all the $P_4$s in $K \cup S$ and since $|T| = \binom{|S|}{2}$, this is a minimum set.
\hfill $\Box$
\end{proof}

\begin{theorem}
 Algorithm~\ref{alg:spider} correctly solves the cograph edge-deletion problem for $P_4$-sparse graphs and can be implemented in $O(m+n)$ time.
\end{theorem}

\begin{proof}
 The correctness of Algorithm~\ref{alg:spider} follows from Lemma~\ref{thin}, Lemma~\ref{thick} and Proposition~\ref{prop:decomposition}.

 Algorithm~\ref{alg:spider} can be implemented in linear time, as the spider structure of $P_4$-sparse graphs can be identified in linear time~\cite{JaOl}. Identifying the connected or co-components can also be done in linear time, as these types of vertex partitions are special cases of the more general notion of a \emph{homogeneous set} or \emph{module}, and there are a number of modular decomposition algorithms running in linear time~\cite{McCSp},~\cite{CoHa}.
\hfill $\Box$
\end{proof}

Our algorithm to find cograph edge-deletion sets in $P_4$-sparse graphs is presented in Algorithm~\ref{alg:spider}.

\begin{algorithm}[H]
\SetAlgoLined Algorithm {\sc Spider($G$)}:\\
\KwIn{A $P_4$-Sparse Graph $G=(V,E)$}
\KwOut{A set $S \subset E$ such that $(V,E\setminus S)$ is a $P_4$-free graph}
\ \\
\If{$G$ (or $\overline{G}$) is disconnected}{
    Let $V_1, \ldots, V_t$ be the components or co-components of $G$\;
    $S \leftarrow S \ \bigcup_{i=1}^t${\sc Spider}($V_i$)\;
}
$G$ is a spider with $K = \{k_1, \ldots , k_{|K|}\}$ and $S = \{s_1,\ldots ,s_{|K|}\}$\;
\If{$G$ is a thin spider}{
    Notation: $k_i$ adjacent to $s_j$ if and only if $i=j$\;
    Add edge $\{k_i,s_i\}$ to solution set $S$ for every $i=1,\ldots ,|K|-1$\;
}
\If{$G$ is a thick spider}{
    Notation: $k_i$ adjacent to $s_j$ if and only if $i\neq j$\;
    Add edge $\{k_i,s_j\}$ to solution set $S$ for every pair $i < j$\;
}
Return $S \ \cup $ {\sc Spider($R$)}\;
\ \\
\caption{Cograph edge-deletion algorithm for $P_4$-sparse graphs}
\label{alg:spider}
\end{algorithm}

\subsection{A Bounded Search Tree Algorithm for Cograph Edge-Deletion}

The bounded search tree algorithm (Algorithm~\ref{alg:cograph}) finds 5-vertex subsets that induce at least 2 $P_4$s, branches on the possible ways of destroying the $P_4$s, and then finally arrives at a $P_4$-sparse graph and calls Algorithm~\ref{alg:spider}. This algorithm either terminates with a call to the subroutine (in the case that a spider structure is encountered) or detects a cograph structure early, or else its integer parameter $k$ has been reduced to 0 or less in which case the number of allowed edge-deletions has been exhausted without reaching a cograph.

Refer to Figure~\ref{forb} for the possible subgraphs the general search algorithm may encounter. We refer to specific edges as they are labeled in Figure~\ref{forb} for each subgraph. The pseudocode description of the general search algorithm branches on one of the deletion sets given in the table below.

Let $H$ be one of the forbidden subgraphs from Figure~\ref{forb}. The possible edge-deletion sets to destroy the $P_4$s in $H$ are:
$$H = \left\{
\begin{tabular}{c|c|}
  \hline
  Subgraph & Minimal Edge Deletion Sets \\ \hline
  $C_5$ & \{a,c\}, \{a,d\}, \{b,d\}, \{b,e\}, \{c,e\}\\
  $P_5$ & \{a,d\}, \{b\}, \{c\}\\
  $\overline{P}_5$ & \{a,b\}, \{e,c\}, \{d,e\}, \{c,d\}, \{a,d,f\}, \{a,c,f\}, \{b,d,f\}, \{b,e,f\}\\
  4-pan & \{a,d\}, \{a,c\}, \{b,c\}, \{b,d\}, \{e\}\\
  co-4-pan & \{b,c\}, \{d\}, \{e\}\\
  fork & \{a,b\}, \{c\}, \{d\}\\
  kite & \{a,d\}, \{a,c,f\}, \{b,d,f\}, \{b,c\}, \{e\}\\
  \hline
\end{tabular}
\right.$$

\ \\

\begin{algorithm}[H]
\SetAlgoLined Algorithm {\sc CographDeletion($G,k$)}\\
\KwIn{A Graph $G=(V,E)$ and a positive integer $k$}
\KwOut{A set $S$ of edges of $G$ with $|S| \leq k$ where $(V,E\setminus S)$ is a cograph if it exists, otherwise {\sc No} }
\ \\
Initialize $S = \emptyset$\;
\If{$G$ is a cograph}{
    Return $S$\;
}
\If{$k \leq 0$}{
    Return {\sc No}\;
}
Apply a $P_4$-sparse recognition algorithm\;
\If{$G$ is $P_4$-sparse}{
    $S \leftarrow S \ \cup $ {\sc Spider($G$)}\;
    If $|S| \leq k$, return $S$; Otherwise, return {\sc No}\;
}
\Else{A forbidden graph $H$ from Figure~\ref{forb} exists\;
    \ForEach{minimal edge-deletion set $E'$ for $H$}{
        $S \leftarrow S \ \cup E'$\;
        {\sc CographDeletion}($G-E',k-|E'|$)\;
    }
}
\ \\
\caption{Bounded search tree algorithm computing a cograph edge-deletion set}
\label{alg:cograph}
\end{algorithm}

It is routine to verify that any edge-deletion set from each of the 7 induced subgraph cases must contain one of the deletion set cases given in the table. Since every $P_4$ in the graph must be destroyed with an edge deletion, encountering any of these 7 configurations necessitates the need to apply one of the corresponding deletions.

The runtime of the algorithm is dominated by the size of the search tree. The spider structure can be identified in linear time. When $k$ is the parameter measuring the number of edge deletions left to make, the size $T(k)$ of the search tree produced by this process is found from each branch rule separately:
\begin{enumerate}
  \item $C_5$: five branches, each reducing the parameter by 2 gives $T(k) = 5T(k-2)$ and so $T(k) \leq 2.237^k$
  \item $P_5$: $T(k) = 2T(k-1) + T(k-2)$ giving $T(k) \leq 2.415^k$
  \item $\overline{P}_5$: $T(k) = 4T(k-2) + 4T(k-3)$ giving $T(k) \leq 2.383^k$
  \item 4-pan: $T(k) = T(k-1) + 4T(k-2)$ giving $T(k) \leq 2.562^k$
  \item co-4-pan: $T(k) = 2T(k-1) + T(k-2)$ giving $T(k) \leq 2.415^k$
  \item fork: $T(k) = 2T(k-1) + T(k-2)$ giving $T(k) \leq 2.415^k$
  \item kite: $T(k) = T(k-1) + 2T(k-2) + 2T(k-3)$ giving $T(k) \leq 2.270^k$
\end{enumerate}

The size of the search tree is thus upper-bounded by the worst case of deleting $P_4$s in a 4-pan: $T(k) \leq 2.562^k$.

\begin{theorem}
Algorithm~\ref{alg:cograph} correctly solves the cograph $k$-edge-deletion problem in {$O(2.562^k(n+m))$} time.
\end{theorem}

\begin{proof}
Jamison and Olariu~\cite{JaOl} give a linear time recognition algorithm for $P_4$-sparse graphs. In the case that the graph being tested is not $P_4$-sparse, the algorithm terminates upon finding a 5-set of vertices isomorphic to one of the forbidden subgraphs shown in Figure~\ref{forb}. In $O(m+n)$ time on a general graph, we can find one of the subgraphs in Figure~\ref{forb} or else assert that our graph is $P_4$-sparse.

\hfill $\Box$
\end{proof}

\subsection{A Bounded Search Tree Algorithm for Edge-Deletion to Trivially Perfect Graphs}

In \cite{Guo}, Guo studied the edge-deletion problem for \emph{complements} of trivially perfect graphs. We know of no prior study of the specific problem of deleting edges to a trivially perfect graph. A na\"ive solution would find a subgraph isomorphic to either a $P_4$ or a $C_4$ and then branch on the possible ways of deleting an edge from that subgraph, resulting in a worst-case search tree of size $O(4^k)$. A minor observation that deleting any one edge from a $C_4$ always results in the other forbidden subgraph, $P_4$, allows us to branch on the 6 possible ways of deleting any 2 edges from a $C_4$. This results in a worst-case search tree of size $O(3^k)$ due to the 3 edges in a $P_4$.

We use our strategy of branching towards a relaxation class of trivially perfect graphs. The 6 possible ways of deleting 2 edges from $C_4$ yield a branching factor of $\sqrt{6}^k \leq 2.45^k$, and since removing two edges from any $C_4$ is necessary to arrive at a ($P_4$,$C_4$)-free graph, our algorithm will begin by performing this branching step before running a $P_4$-sparse recognition algorithm. Then we proceed as in the previous section, finding any $P_4$-sparse forbidden subgraph and branching on the ways of deleting $P_4$s and $C_4$s in it. Once no $P_4$-sparse obstruction exists, we solve the problem optimally on the resulting specialized structure (a $C_4$-free $P_4$-sparse graph.) The branching rules become simpler in that only 5 of the 7 graphs in Figure~\ref{forb} need consideration. In particular, the \emph{4-pan} that caused the bottleneck of Algorithm~\ref{alg:cograph}, is no longer considered and this changes the runtime of the process from $O(2.562^k)$ to $O(2.450^k)$.

One main difference in this algorithm from Algorithm~\ref{alg:cograph} is that $C_4$s are found and destroyed first, and after any of the $P_4$-sparse deletions are made, the process restarts with looking for $C_4$s to destroy again. Once the $C_4$s are destroyed and the resulting graph is $P_4$-sparse, we proceed with removing edges with edge-deletion algorithm for thin or thick spiders (Algorithm~\ref{alg:spider}).

\begin{algorithm}[H]
\SetAlgoLined Algorithm {\sc TriviallyPerfectEdgeDeletion($G,k$)}\\
\KwIn{A Graph $G=(V,E)$ and a positive integer $k$}
\KwOut{A set $S$ of edges of $G$ with $|S| \leq k$ where $(V,E\setminus S)$ is trivially perfect if it exists, otherwise {\sc No} }
\ \\
Initialize $S = \emptyset$\;
\If{$G$ is a trivially perfect}{
    Return $S$\;
}
\If{$k \leq 0$}{
    Return {\sc No}\;
}
\While{There exists $H$ isomorphic to $C_4$}{
    Create 6 branches corresponding to the possible ways of removing any 2 edges in $H$
}
Apply a $P_4$-sparse recognition algorithm\;
\If{$G$ is $P_4$-sparse}{
    $S \leftarrow S \cup $ {\sc Spider($G$)}\;
    If $|S| \leq k$, return $S$; Otherwise, return {\sc No}.
}
\Else{A forbidden graph $H$ from Figure~\ref{forb} exists\;
    \ForEach{minimal vertex-deletion set $S'$ for $H$}{
        Add the vertices $S'$ to the solution set $S$\;
        {\sc TriviallyPerfectEdgeDeletion}($G-S'$, $k - |S'|$)\;
    }
}
\ \\
\caption{Bounded search tree algorithm finding a trivially perfect edge-deletion set}
\label{alg:TrivPer}
\end{algorithm}

The correctness of decomposing the edge-deletion problem into separate problems on $K\cup S$ and $R$ depends a proposition similar to Proposition~\ref{prop:decomposition}.

\begin{proposition}\label{prop:trivperdecomp}
 Let $G$ be a $C_4$-free graph and $M(G)$ be the size of a minimum edge-deletion set required to turn $G$ into a $(P_4,C_4)$-free graph. Then:
 \begin{itemize}
 \item[i)] if $G$ is disconnected with components $V_1, \ldots , V_t$, then $M(G) = \sum_{i=1}^t M(V_i)$
 \item[ii)] if $\overline{G}$ is disconnected, $G$ is a complete join between a clique and a smaller $C_4$-free graph, $H$, and $M(G) = M(H)$.
 \item[iii)] if $G$ is a spider with head $R$, body $K$ and feet $S$, then $$M(R\cup K\cup S) = M(R) + M(K\cup S)$$.
 \end{itemize}
\end{proposition}

\begin{proof}

 Case i): If $G$ has more than one connected component, any edge deletions made in one component cannot create a $P_4$ or a $C_4$ in a different connected component.

 Case ii): $\overline{G}$ is disconnected. Let $H$ be a set of at least 2 vertices inducing a connected component in $\overline{G}$. Then $H$ induces a $C_4$-free graph in $G$ since any induced subgraph of a $C_4$-free graph is $C_4$-free. Since $H$ is connected in $\overline{G}$, there must be two non-adjacent vertices $u, v$ of $H$ in $G$. Let $x$ and $y$ be any two vertices not in $H$. If $x$ and $y$ are not adjacent, then $\{u,x,y,v\}$ induces a $C_4$ in $G$, which is impossible. So any vertices outside of $H$ must induce a clique in $G$. Furthermore, since $H$ is a connected component in $\overline{G}$, every vertex in $H$ is adjacent to every vertex of the clique $G\setminus H$. It follows, then, that no $P_4$ in $G$ includes a vertex of $G\setminus H$, and after any edge deletions in $H$, no $P_4$ or $C_4$ can include a vertex of $G\setminus H$. Hence $M(G)=M(H)$.

 Case iii): Notice that no $C_4$ can include a vertex $s$ from $S$ in a spider even after removals of leg edges and head edges since the neighbourhood of $s$ induces a clique. Since $K$ is a clique, and every $k \in K$ is adjacent to every $r \in R$, there can not exist a $C_4$ in $K \cup R$ unless the $C_4$ is completely contained in $R$. So no $C_4$ contains an edge from $R$ to $K$. Therefore, any edge $e = \{r,k\}$ with $r \in R$ and $k \in K$ is not in any $C_4$ in $G$, and for any subset of leg edges and head edges $E'$ the edge $e = \{r,k\}$ is not in any $C_4$ in $G-E'$. Combining this with Proposition~\ref{prop:decomposition} for $P_4$s establishes the decomposition.
\hfill $\Box$

\end{proof}

Proposition~\ref{prop:trivperdecomp} shows us that since all the $C_4$s are destroyed in the branching stage of {\sc TriviallyPerfectEdgeDeletion($G,k$)}, once we arrive at a $C_4$-free spider, we are free to delete leg edges without creating a new $C_4$.




The runtime of Algorithm~\ref{alg:TrivPer} is dominated by the branching rules once again. Encountering a $C_4$ results in 6 branches which delete 2 edges each. The resulting recurrence is $T(k) = 6T(k-2)$ and so $T(k) \leq 2.450^k$. Having deleted all the $C_4$s, we no longer include the $\overline{P}_5$ or the \emph{4-pan} cases in our analysis. The runtime analysis for the rest remain unchanged: $C_5: 2.237^k, P_5: 2.415^k$, co-4-pan: $2.415^k$, fork: $2.415^k$, kite: $2.270^k$. The search tree is thus bounded by the $C_4$ case of size $O(2.450^k)$. Finding a $C_4$ directly is a problem that is currently best-achieved using matrix multiplication~\cite{KrSp}, so this entire process \emph{as described} runs in $O(2.450^kn^\alpha)$ where $O(n^\alpha)$ is the time required for matrix multiplication ($\alpha \leq 2.376$~\cite{CopWin}).

We can, in fact, modify the algorithm to run linearly in $n$ and $m$ by observing that a graph is $P_4$-free and $C_4$-free if and only if it is a chordal cograph. By first running a certifying chordal recognition algorithm~\cite{TaYa}, we can either deduce that there is no $C_4$ or else find a $C_4$ or a $C_5$ or a larger induced cycle (and thus a $P_5$) and branch on these subgraphs according to the rules we gave, and if the graph is chordal then we apply a $P_4$-sparse recognition algorithm to find one of the other forbidden induced subgraph, branch on it, and then re-apply the chordal recognition process.

\begin{theorem}
Finding a trivially perfect $k$-edge-deletion set can be solved in $O(2.450^k(n+m))$ time.
\end{theorem}

\section{Vertex-Deletion Algorithms}

This section shows how our general method can be used to solve vertex-deletion version of our prior two problems:

\begin{problem}{\sc Cograph Vertex-Deletion} $(G, k)$:\\
Given graph $G=(V,E)$, does there exist a set $S$ of at most $k$ vertices such that $G-S$ is a cograph?
\end{problem}

\begin{problem}{\sc Trivially Perfect Vertex-Deletion} $(G, k)$:\\
Given graph $G=(V,E)$, does there exist a set $S$ of at most $k$ vertices such that $G-S$ is a trivially perfect graph?
\end{problem}

\subsection{Vertex-Deletion to Cographs}

Since removing a vertex set $S$ from a graph $G=(V,E)$ is equivalent to taking the induced subgraph on the vertex set $V\setminus S$, these problems are also often named \emph{maximum induced subgraph} problems. In our case of asking if there is a vertex set of size at most $k$ that can be removed to leave behind a cograph, this is equivalent to asking if there is an induced cograph subgraph of size at least $|V|-k$. Removing a vertex from $G$ can never create a new induced subgraph in $G$, and so deleting vertices to destroy induced subgraphs is commonly modeled as a {\sc Hitting Set} problem. In this case in which each $P_4$ maps to a 4-set in a {\sc Hitting Set} instance, we have the restricted problem of a 4-{\sc Hitting Set}. Algorithms for such vertex-deletion problems should always be compared against the state-of-the-art algorithms of $d-${\sc Hitting Set} if not anything else. $d$-{\sc Hitting Set} is a well-studied NP-complete problem which admits fixed-parameter tractable algorithms. The first improved analysis of $d$-{\sc Hitting Set} by Niedermeier and Rossmanith~\cite{NiRo} give a search tree of size $O(3.30^k)$, and a more detailed and involved analysis by Fernau~\cite{Fer} improves the bound to $O(3.148^k)$. This is the best known bound for $4$-{\sc Hitting Set} to date.

The simple spider structure of $P_4$ sparse graphs allows us to describe a simple algorithm for the vertex-deletion problem to cographs. The runtime of this simple algorithm matches that of~\cite{GGHN} and of~\cite{NiRo}. The algorithm in~\cite{GGHN} used branching rules that were designed by breaking the $P_4$s in every subgraph of size $t$. Testing various values of $t$ deduced that rules based on subgraphs of size 7 yielded the optimal runtime of an algorithm of this sort, with runtime $O(3.30^k)$. Their algorithm builds branching rules from 447 graphs of size 7, while our algorithm only involves seven graphs on 5 vertices (Figure~\ref{forb}.)

In the following subsection, we use the analysis technique of~\cite{NiRo} to show that the runtime of our bounded search tree algorithm is $O(3.115^k)$, hence improving on Fernau's $O(3.148^k)$. Our runtime could be improved further if we were to use the methods of Fernau~\cite{Fer}, but such an analysis is extensive and would sidetrack from the focus of this paper.

We describe the subroutine {\sc Spider Vertex-Deletion} here. The algorithm works in the same way as Algorithm~\ref{alg:spider}, taking as input a $P_4$-sparse graph and returning the optimal number of vertices to remove in order to break all $P_4$s in the graph. For thin spiders, every pair of feet is the end-pair of a $P_4$, and removing any $|S|-1$ vertices from $S$ will destroy all the $P_4$s in the body and legs. Removing less than $|S|-1$ will leave at least two thin legs and hence a $P_4$, so $|S|-1$ is necessary.

Since a set of 4 vertices induces a $P_4$ in a graph $G$ if and only if they induce a $P_4$ in $\overline{G}$, deleting any $|K|-1$ vertices from $K$ in a thick spider will destroy all the $P_4$s in $K \cup S$. In either the thin or thick spider case, the subroutine is then applied to head $R$. This concludes the description of {\sc Spider Vertex-Deletion}. The correctness of the algorithm is asserted by the following proposition:

\begin{proposition}\label{prop:cographvertex}
Let $G$ be a spider with head $R$, body $K$ and feet $S$. Let $M(G)$ be the minimum number of vertices required to remove from $G$ in order to turn $G$ into a cograph. Then $M(G) = M(R) + M(S \cup K)$.
\end{proposition}

\begin{proof}
  Deleting vertices from a graph can never create a new $P_4$. We know from Proposition~\ref{prop:decomposition} that no $P_4$ includes vertices from both $K$ and $R$. Deleting any vertices from $K\cup S$ will not destroy $P_4$s in $R$, and vertex deletions from $R$ will not destroy any $P_4$s in $K\cup S$. Hence $M(G) = M(R) + M(S \cup K)$.
  \hfill $\Box$
\end{proof}

\begin{corollary}
  The algorithm {\sc Spider Vertex-Deletion} described above correctly solves the cograph vertex deletion problem for spiders in linear time.
\end{corollary}

For a general graph, we proceed as in the cograph edge-deletion algorithm. We find $P_4$-sparse obstructions and branch on the possible ways of deleting vertices to destroy all $P_4$s, repeating until the remaining graph is $P_4$-sparse. The pseudocode description is given in~\ref{alg:cographVertex}.

\begin{algorithm}[H]
\SetAlgoLined Algorithm {\sc CographVertexDeletion($G,k$)}\\
\KwIn{A Graph $G=(V,E)$ and a positive integer $k$}
\KwOut{A set $S$ of vertices of $G$ with $|S| \leq k$ where $(V\setminus S,E)$ is a cograph if it exists, otherwise {\sc No} }
\ \\
Initialize $S = \emptyset$\;
\If{$G$ is a cograph}{
    Return $S$\;
}
\If{$k \leq 0$}{
    Return {\sc No}\;
}
Apply a $P_4$-sparse recognition algorithm\;
\If{$G$ is $P_4$-sparse}{
    $S \leftarrow S \cup $ {\sc Spider Vertex-Deletion($G$)}\;
    If $|S| \leq k$, return $S$; Otherwise, return {\sc No}.
}
\Else{A forbidden graph $H$ from Figure~\ref{forb} exists\;
    \ForEach{minimal vertex-deletion set $S'$ for $H$}{
        Add the vertices $S'$ to the solution set $S$\;
        {\sc CographVertexDeletion}($G-S'$, $k-|S'|$)\;
    }
}
\ \\
\caption{Bounded search tree algorithm finding a cograph vertex-deletion set}
\label{alg:cographVertex}
\end{algorithm}

The branching rules for the vertex deletions are given in a table as before:

$$H = \left\{
\begin{tabular}{c|c|}
  \hline
  Subgraph & Minimal Vertex Deletion Sets \\ \hline
  $C_5$ & \{1,2\}, \{1,3\}, \{1,4\}, \{1,5\}, \{2,3\}, \{2,4\}, \{2,5\}, \{3,4\}, \{3,5\}, \{4,5\}\\
  $P_5$ & \{1,5\}, \{2\}, \{3\}, \{4\}\\
  $\overline{P}_5$ & \{1\}, \{3\}, \{4\}, \{2,5\}\\
  4-pan & \{2\}, \{4\}, \{5\}, \{1,3\}\\
  co-4-pan & \{3\}, \{4\}, \{5\}, \{1,2\}\\
  fork & \{3\}, \{4\}, \{5\}, \{1,2\}\\
  kite & \{2\}, \{4\}, \{5\}, \{1,3\}\\
  \hline
\end{tabular}
\right.$$

The runtime of the algorithm is dominated by the branching steps. The runtime $T(k)$ for the $C_5$ case depends on 10 branches, while each of the other cases have equivalent runtime analysis.

\begin{enumerate}
  \item $C_5$: ten branches, each reducing the parameter by 2 gives $T(k) = 10T(k-2)$ and so $T(k) \leq 3.163^k$
  \item All others: $T(k) = 3T(k-1) + T(k-2)$ giving $T(k) \leq 3.303^k$
\end{enumerate}

The runtime of this vertex-deletion algorithm is bounded by $O(3.303^k(m+n))$, matching the runtime of the cograph vertex deletion algorithm generated by automated branching rule design~\cite{GGHN}.

\begin{theorem}
    Algorithm~\ref{alg:cographVertex} solves the vertex-deletion problem for cographs in $O(3.303^k(m+n))$ time.
\end{theorem}

\subsection{Improvement using Hitting-Set}

The 4-{\sc Hitting Set} algorithm of~\cite{NiRo} involves an analysis which counts when a branch choice can be made on a 3-set. Without counting these cases, an algorithm for 4-{\sc Hitting Set} which only makes choices on 4-sets will have a search tree size of $4^k$. By keeping track of when 3-sets are created in the search process and by branching on 3-sets whenever they are available, the authors of~\cite{NiRo} are able to improve the upper-bound to the size of the search tree to $O(3.30^k)$.

We show here that using a similar technique to that in~\cite{NiRo} can improve our search tree size. For an instance $(G,k)$ of cograph vertex-deletion, we will use an implicit instance of 4-{\sc Hitting Set} where each set of 4 vertices inducing a $P_4$ in $G$ corresponds to a 4-set. A cograph deletion set for $G$ will correspond to the hitting set of the set of 4-sets. 

We adapt the notion of dominance from {\sc Hitting Set} to that of $P_4$s: a vertex $v$ \emph{$P_4$-dominates} $u$ if $v$ exists in every $P_4$ that $u$ is in.

Following the {\sc Hitting Set} method, we observe that if $v$ $P_4$-dominates $u$, then any hitting set that contains $u$ could be replaced with a hitting set containing $v$. Using this observation, we \emph{mark} $u$ in the graph to signify that it will not be in our solution. When we encounter $P_4$s involving $u$, the $P_4$ only needs to be considered as a 3-set to hit. Marking $u$ in $G$ is equivalent to removing $u$ in the implicit 4-{\sc Hitting Set} instance.

Our vertex-deletion algorithm given in the previous subsection applies a $P_4$-sparse recognition algorithm to find one of the 7 forbidden configurations from Figure~1. We illustrate how to proceed to the branching step when encountering a $P_5$: $\{v_1,v_2,v_3,v_4,v_5\}$ with $v_1$ and $v_5$ as the endpoints:

If we put $v_2$ in $S$, remove $v_2$ from the graph $G$ and reduce the parameter $k$ by 1. Any set in the hitting set instance $H$ containing $v_2$ is removed. Otherwise (if $v_2$ is not in $S$) we mark $v_2$ in $G$ and remove $v_2$ from $H$, possibly creating some 3-sets. If $v_3$ is put in $S$, reduce $k$ by 1 and remove $v_3$ from the graph, as before. Otherwise (if $v_3$ is also not in $S$) then mark $v_3$ in the graph and remove $v_3$ from $H$. If $v_4$ is in $S$, reduce $k$ by 1 and remove any set containing $v_4$. Otherwise (if none of $v_2, v_3, v_4$ are in $S$) we add $v_1$ and $v_5$ to $S$, remove them from $G$ and reduce the parameter $k$ by 2.

If we first ensure that $P_4$-dominated vertices have been removed from consideration, some vertices in the $P_5$ (or analogous forbidden subgraph) may be marked. We do not need to build branches on cases asking if a vertex $v$ is in $S$ if $v$ is already marked. If we encounter a $P_4$ in one of our forbidden subgraphs consisting of four marked vertices, we can terminate that branch of the search tree and backtrack.

When encountering any of the $P_4$-sparse forbidden subgraphs: $P_5, \overline{P}_5$, kite, fork, 4-pan, co-4-pan, we have in each case 3 vertices whose removal will break both $P_4$s in the obstruction, or else two vertices which must be removed together. Call those first 3 vertices the \emph{breaking vertices.} Our process is summarized in the following algorithm:\\

\begin{algorithm}[H]
\SetAlgoLined Algorithm {\sc CographVertexDeletionHittingSet($G,k$)}\\
\KwIn{A Graph $G=(V,E)$ and a positive integer $k$}
\ \\
1. If any 3-set has been created, branch on that 3-set. Repeat until there are no more 3-sets\;
2. If any vertex is $P_4$-dominated, mark it in $G$ and remove it from the hitting set. Go to step 1.\;
3. Find one of $P_5, \overline{P}_5$, kite, fork, 4-pan, co-4-pan.\;
4. If there is a $P_4$ all of whose vertices are marked, {\sc Stop} and backtrack.\;
5. Branch on the (up to 3) unmarked breaking vertices using the cases as described above. Go to step 1.\;
6. If all three breaking vertices are marked, include the other two vertices in $S$ and Go to 1.\;
7. If no such subgraph can be found, our graph is an \emph{extended $P_4$-sparse graph} (See below.) Solve the remainder optimally without search.\\
\ \\
\caption{Using Hitting-Set for Cograph Vertex-Deletion}
\label{alg:cographVertexHittingSet}
\end{algorithm}

Let $v$ and $v'$ be breaking vertices encountered after steps 1 and 2 cannot be applied any further. If $v$ is not in any other $P_4$ besides the two $P_4$s in the obstruction graph found in step 3, then $v$ is $P_4$-dominated by $v'$, but this cannot happen if steps 1 and 2 are done to exhaustion. So we have that $v$ must be in another $P_4$ not involving $v'$. When branching on $v$, we consider $v \in S$, in which case $v$ is removed from $G$, or $v \notin S$ in which case we remove $v$ from $H$, creating at least one 3-set since we established that $v$ must be in another $P_4$ not containing $v'$.

Step 3 can be performed with a linear-time algorithm recognizing ($P_5, \overline{P}_5$, kite, fork, 4-pan, co-4-pan)-free graphs. These are called \emph{extended $P_4$-sparse graphs} by Giakoumakis and Vanherpe~\cite{GV}. This ensures we do not encounter a $C_5$ at this stage of the process. They showed:

\begin{theorem}\cite{GV}
If $C$ is a $C_5$ in an extended $P_4$-sparse graph, then $C$ is a prime module.
\end{theorem}

\begin{figure}
    \hspace{3.6cm}
  \includegraphics[height=2cm]{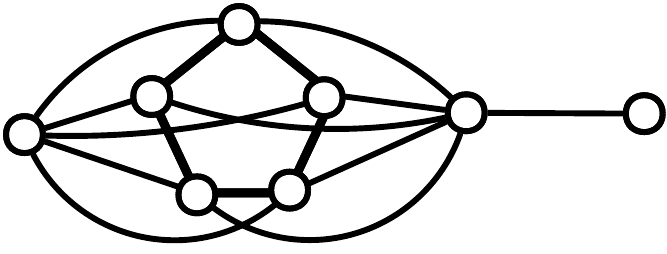}\\
  \caption{An impossible configuration for a $C_5$ in an extended $P_4$-sparse graph}\label{impossible}
\end{figure}

Let $C$ be a $C_5$ in our graph after reaching step 7 of our hitting-set process. Observe that every 4-set of $C$ induces a $P_4$, so no vertex of $C$ is contained in a nontrivial module or else we will have one of the forbidden subgraphs of extended $P_4$-sparse graphs which we have already destroyed. Further, $C$ can not be a module in some $P_4$ or else that $P_4$ extends to one of the forbidden graphs already destroyed (see Figure~\ref{impossible}.) It must be that $C$ is a set of 5 vertices inducing a 5-cycle and not overlapping with any other $P_4$. Since $C$ does not intersect with any other existing $P_4$s left in $G$, we are free to choose any two vertices of $C$ to add to $S$ and delete from $G$.

After deleting every $C_5$ from the extended $P_4$-sparse graph, we have a conventional $P_4$-sparse graph and we proceed with vertex deletions for spiders using {\sc Spider Vertex-Deletion} described in the previous subsection.

Let $T(k)$ be the number of leaves in a search tree of our vertex deletion problem, and let $B(k)$ be the number of leaves in a search tree for this problem whose root branched on a 3-set. Step 4 of Algorithm~\ref{alg:cographVertexHittingSet}, can (at worst) branch on each of the breaking vertices. If the first vertex is put in $S$, we reduce $k$ by 1 and so we have a $T(k-1)$ branch. If we assume the first vertex is not in $S$ and select the second vertex to be in $S$, the parameter decreases by 1. Since this first vertex is not $P_4$-dominated (or else it would have been marked), it must be in another $P_4$ and so marking the it will create at least one 3-set in $H$, giving a $B(k-1)$ branch. Along the same lines, if we choose the third breaking vertex, we arrive at another $B(k-1)$ branch. In the final case of deleting the two non-breaking vertices, we create a $T(k-2)$ branch. Together this puts an upper bound on $T(k)$ of $T(k-1) + 2B(k-1) + T(k-2)$.

Similarly, when branching on a 3-set, $B(k) \leq T(k-1) + 2B(k-1)$. Together, these two recurrences give a simultaneous system from which one can show $B(k) \leq 3.115^k$ and $T(k) \leq 1.115B(k)$ with a straightforward induction proof.

\begin{theorem} Algorithm~\ref{alg:cographVertexHittingSet} solves the cograph vertex-deletion problem in $O(3.115^k)$ time.
\end{theorem}

The method of analyzing the search tree size created upon the existence of a 3-set shows that our search tree size is smaller than the $O(3.30^k)$ for 4-{\sc Hitting Set} found by Niedermeier and Rossmanith~\cite{NiRo}. Fernau~\cite{Fer} refines this analysis process by keeping track of the \emph{the number} of $(d-1)$-sets in $d$-{\sc Hitting Set}, arriving at $O(3.148^k)$ for 4-hitting set. Specifically, Fernau's analysis involves expressions $T^i(k)$ for $i=0,1,2,3$ where $i$ is the number of 3-sets in an instance of 4-hitting set (in our case, $B(k)$ is $T^1(k)$.) We are confident that a similar refinement in the analysis of our vertex-deletion algorithm would reveal further gains, but our presented algorithm is already shown to have a smaller search space.

\subsection{Vertex-Deletion for Trivially Perfect Graphs}

Given a graph $G$, our task now is to find the largest induced trivially perfect subgraph in $G$. Equivalently, given a value $k$, we want know whether we can delete at most $k$ vertices in order to turn the graph $P_4$-free and $C_4$-free.

In the edge-deletion version of this problem from the previous section, we deleted at least 2 edges from all $C_4$s in the branching process since 2 edges is necessary, and this was algorithmically appealing as it decreased the parameter by 2. The vertex-deletion problem does not share this luxury: there are 4 vertices in a $C_4$ and only a single vertex removal is required to turn it into a ($P_4$, $C_4$)-free graph. This will result in a more complicated procedure to delete all remaining $C_4$s in the $P_4$-sparse graph that remains after the search process.

We will proceed directly to finding the $P_4$-sparse obstructions and branching on them to turn each one into a $(P_4, C_4)$-free graph. This yields a worst-case runtime of $O(3.303^k)$, as summarized by the following table for each obstruction graph $H$:

$$H = \left\{
\begin{tabular}{c|c|}
  \hline
  Subgraph & Minimal Vertex Deletion Sets \\ \hline
  $C_5$ & \{1,2\}, \{1,3\}, \{1,4\}, \{1,5\}, \{2,3\}, \{2,4\}, \{2,5\}, \{3,4\}, \{3,5\}, \{4,5\}\\
  $P_5$ & \{1,5\}, \{2\}, \{3\}, \{4\}\\
  $\overline{P}_5$ & \{1,2\}, \{1,3\}, \{1,4\}, \{1,5\}, \{2,3\}, \{2,4\}, \{2,5\}, \{3,4\}, \{3,5\}, \{4,5\}\\
  4-pan & \{2\}, \{4\}, \{1,3\}, \{1,5\}, \{3,5\}\\
  co-4-pan & \{3\}, \{4\}, \{5\}, \{1,2\}\\
  fork & \{3\}, \{4\}, \{5\}, \{1,2\}\\
  kite & \{2\}, \{4\}, \{5\}, \{1,3\}\\
  \hline
\end{tabular}
\right.$$

The runtime for each of these cases is summarized below:

\begin{enumerate}
  \item $C_5$: five branches, each reducing the parameter by 2 gives $T(k) = 10T(k-2)$ and so $T(k) \leq 3.167^k$
  \item $P_5$: $T(k) = 3T(k-1) + T(k-2)$ giving $T(k) \leq 3.303^k$
  \item $\overline{P}_5$: $T(k) = 10T(k-2)$ giving $T(k) \leq 3.167^k$
  \item 4-pan: $T(k) = 2T(k-1) + 3T(k-2)$ giving $T(k) \leq 3^k$
  \item co-4-pan: $T(k) = 3T(k-1) + T(k-2)$ giving $T(k) \leq 3.303^k$
  \item fork: $T(k) = 3T(k-1) + T(k-2)$ giving $T(k) \leq 3.303^k$
  \item kite: $T(k) = 3T(k-1) + T(k-2)$ giving $T(k) \leq 3.303^k$
\end{enumerate}

After all the forbidden configurations of $P_4$-sparse graphs have been destroyed, we are left with a $P_4$-sparse graph from which we must delete vertices to destroy the remainder of the $P_4$s and $C_4$s. While $C_4$s do not exist in a thin or thick spider, $C_4$s will exist across co-components. Namely, if $A_1$ and $A_2$ are two non-clique co-components, then any two nonadjacent vertices $x_1$ and $y_1$ in $A_1$ and any two nonadjacent vertices $x_2$ and $y_2$ in $A_2$ will induce a 4-cycle. Since each connected component and co-component of a $P_4$-sparse graph must be a spider, every induced 4-cycle must be the type that crosses non-clique co-components.

In order for this $P_4$-sparse graph to be $C_4$ free, all but one of the co-components must be a clique. The only $P_4$s that will be left to delete will be those strictly in the non-clique co-component. To determine the optimal way at arriving at this point, let us introduce some notation: for a $P_4$-sparse graph $G$, let $A_1, A_2, \ldots , A_t$ be the co-components of $G$. Let $\omega_i = \omega(A_i)$ be the size of a maximum clique in $A_i$, and $\eta_i$ be the size of a minimum cograph vertex-deletion set, as found by the algorithm {\sc Spider Vertex-Deletion}.

We seek to find $i$ such that deleting all co-components $A_j, j\neq i$ into cliques, plus {\sc Spider Vertex-Deletion}($A_i$) is a minimum. That is, we want to find $i$ that minimizes
$$\eta_i + \sum_{j\neq i} |A_j|-\omega_j.$$

For a particular $G$, $\sum{|A_i|}=n$ is fixed, as is $\sum{\omega_i}$. We see that the expression above is minimized when $i$ is chosen such that $|A_i| - \omega_i - \eta_i$ is a maximum.

Our algorithm is as follows:

\begin{algorithm}[H]
\SetAlgoLined Algorithm {\sc TriviallyPerfectVertexDeletion($G,k$)}\\
\KwIn{A Graph $G=(V,E)$ and a positive integer $k$}
\KwOut{{\sc Yes} if there exists a set $S$ of at most $k$ vertices so that $G-S$ is trivially perfect, {\sc No} otherwise.}
\ \\
\While{$G$ is not $P_4$-sparse}{
    Let $H$ be a $P_4$-sparse obstruction subgraph\;
    Branch on the possible ways of deleting the $P_4$s and $C_4$s from H\;
    Let $k_1$ be the number of vertex deletions made in this stage\;
}
$G$ is $P_4$-sparse. Let $A_1, \ldots, A_t$ be the co-components of $G$\;
Fix $i$ to be the lowest index maximizing $|A_i| - \omega_i - \eta_i$\;
\For{$j \neq i$}{
    Fix a maximum clique of $A_j$\;
    Delete any vertex of $A_j$ which is not in this maximum clique\;
}
Let $k_2$ be the number of vertex deletions made in the for-loop\;
Let $k_3$ be the number of deletions performed in {\sc Spider Vertex-Deletion}($A_i$)\;
\If{$k_1+k_2+k_3 \leq k$}{
    return {\sc Yes}\;
}
return {\sc No}\;
\ \\
\caption{Trivially Perfect Vertex-Deletion Algorithm}
\label{alg:TrivPerfVertex}
\end{algorithm}

Since maximum cliques can be found in linear time on $P_4$-sparse graphs, it is clear that this algorithm runs in polynomial time for any fixed $k$. The runtime is dominated by the exponential factor from the search tree, which was shown to be $O(3.303^k)$.

\begin{theorem}
Algorithm~\ref{alg:TrivPerfVertex} is a fixed-parameter tractable algorithm which solves the vertex-deletion problem for trivially perfect graphs in $O(3.303^k)$.
\end{theorem}

Of course, a hitting set-style improvement similar to the previous section could be applied here.

\section{Conclusions and Future Work}

We presented a general method for solving a variety of graph modification problems by limiting a search to a superclass of graphs which are close enough to the target that an optimal polynomial-time subroutine can be used. The algorithms presented here depend on the fact that deleting edges or vertices to a cograph or a trivially-perfect graph can be solved in linear time when the input is a $P_4$-sparse graph. For general input, we gave the first non-trivial algorithm for the cograph edge-deletion problem (running in $O(2.562^k)$) and trivially-perfect edge-deletion problem (running in $O(2.450^k)$ time.) We gave simple algorithms to find minimum vertex-deletion sets to cographs and trivially perfect graphs whose runtime of $O(3.303^k)$ matched the existing literature. We also illustrated how a careful branching strategy and refined analysis technique improved the runtime to $O(3.115^k)$ for the cograph vertex deletion problem, and noted that these vertex-deletion problems can likely yield to further improvements with a more detailed analysis.

Our general method of branching toward a superclass of the target class benefits from few branching rules. This paper concentrated on only 7 graph configurations of $P_4$-sparse graphs, allowing to explicitly define the small number branching rules when compared to the automated method of~\cite{GGHN}.

This paper leaves open many opportunities for future work. Firstly, the bound of $O(2.562^k)$ for cograph edge-deletion is due to the basic branching rules when encountering the 4-pan subgraph. Any improvement on this branching step would reduce the bottleneck of the cograph edge-deletion algorithm. Alternately, one could remove the 4-pan from the list of subgraphs to branch on provided this is accompanied with an efficient algorithm to optimally solve cograph edge-deletion on a ($C_5$, $P_5, \overline{P}_5$, kite, fork, co-4-pan)-free graphs (which are a restricted version of \emph{semi $P_4$-sparse graphs}~\cite{FG}. Such an algorithm would improve cograph edge-deletion from $O(2.562^k)$ to $O(2.415^k)$.

The literature on graph classes is extensive~\cite{BLS}, and many of these classes admit polynomial time solutions to many NP-complete problems. We expect that new and fast fixed-parameter tractable algorithms will soon develop through the use of superclasses as we have used in this paper. It would be interesting to see if the linear time algorithms for treewidth and minimum fill-in on permutation graphs~\cite{Mei} and on distance hereditary graphs~\cite{BDK}, or polynomial time algorithms on weakly chordal graphs~\cite{BT99} could be used to design simple FPT algorithms for those problems.

\bibliography{mybibtex}

\end{document}